\documentclass[a4paper,11pt]{article}
\usepackage{pos}
\usepackage{tikz}
\usepackage{tikz-feynman}
\usepackage{braket}

\title{The search for new physics in $B \to K \ell^+\ell^-$ and $B \to K \nu\bar{\nu}$ using precise lattice QCD form factors}
\ShortTitle{The search for new physics in $B \to K \ell^+\ell^-$ and $B \to K \nu\bar{\nu}$}

\author*[a]{W. G. Parrott}
\author[a]{C. Bouchard}
\author[a]{C. T. H. Davies}

\affiliation[a]{SUPA, School of Physics and Astronomy, University of Glasgow, Glasgow, G12 8QQ, UK}

\emailAdd{parrott@yorku.ca}

\abstract{We present HPQCD's improved scalar, vector and tensor form factors for $B \to K$ semileptonic decays, using the heavy-HISQ formalism for more accurate normalisation of the weak currents. Working with masses close to the physical $b$ on the finest ensemble and including three ensembles with physical light quarks, we cover the full physical $q^2$ range with good precision. Our uncertainties at $q^2=0$ are a factor of three better than earlier work.

We compare Standard Model observables using our form factors to experimental measurements for the rare flavour changing neutral current processes $B \to K \ell^+\ell^-$ and $B \to K \nu\bar{\nu}$ and discuss the significance of the tensions that arise.}

\FullConference{%
The 39th International Symposium on Lattice Field Theory,\\
8th-13th August, 2022,\\
Rheinische Friedrich-Wilhelms-Universität Bonn, Bonn, Germany
}

\begin{document}
\maketitle

\section{Introduction}
\begin{figure}
  \begin{center}
    \begin{tikzpicture}
      \begin{feynman}
        \vertex (a1) {\(\overline b\)};
        \vertex[right=1.5cm of a1] (a2);
        \vertex[right=1.0cm of a2] (a3);
        \vertex[right=1.0cm of a3] (a4);
        \vertex[right=1.5cm of a4] (a5) {\(\overline s\)};
        \vertex[below=1.0cm of a3] (a6);

        \vertex[above=1cm of a1] (b1) {\(u\)};
        \vertex[above=1cm of a5] (b2) {\(u\)};
        \vertex[below=1.5cm of a5] (c1) {\(\ell^-\)};
        \vertex[below=2.0cm of a5] (c3) {\(\ell^+\)};
        \vertex at ($(c1)!0.5!(c3) - (1cm, 0)$) (c2);
        \diagram* {
        {[edges=fermion]
        (a5) -- (a4),
        },
        (b1) -- [fermion] (b2),
        (a2) -- [fermion] (a1),
        (a6) -- [fermion, bend left, edge label=\(\bar{t}\)] (a2),
        (a4) -- [fermion,bend left, edge label=\(\bar{t}\)] (a6),
        (c3) -- [fermion, out=180, in=-60] (c2) -- [fermion, out=60, in=180] (c1),
        (a6) -- [photon, bend right] (c2),
        (a2) -- [boson, edge label=\(W^{+}\)] (a4)
        };
      \end{feynman}
    \end{tikzpicture}
  \end{center}
  \caption{Feynman diagram for a $B^+\to K^+\ell^+\ell^-$ decay.}
  \label{fig:BKfeyn}
\end{figure}
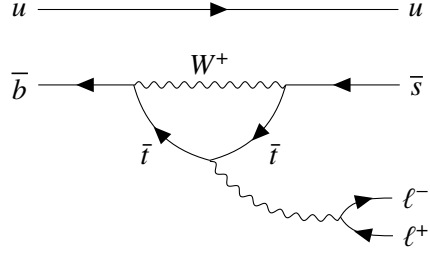
\begin{table}
  \caption{Gluon field ensembles used in this work, numbered in column 1. Column 2 gives the approximate value of $a$ for each set. Column 3 gives the spatial ($N_x$) and temporal ($N_t$) dimensions of each lattice in lattice units and column 4, the number of configurations and time sources used in each case. Columns 5-7 give the masses of the valence and sea quarks in lattice units, noting that $m_u=m_d=m_l$ and the valence and sea masses are the same in the case of $m_l$. Column 8 gives the heavy masses on each ensemble. In each case, the lowest mass is tuned to that of the charm. The sea charm mass is not included here, but is close to the valence value. For more details, see~\cite{BK}}
  \begin{center} 
    \begin{tabular}{ c c c c c c c c}
      \hline
      Set & $a$(fm) & $N_x^3\times N_t$  &$n_{\mathrm{cfg}}\times n_{\mathrm{src}}$ &    $am_{l}^{\mathrm{sea/val}}$ & $am_{s}^{\mathrm{sea}}$ & $am_{s}^{\mathrm{val}}$ & $am_h$\\
  \hline
      \hline
    1    & 0.15 &  $32^3\times 48$    & $998\times 16$ &     0.00235        &  0.0647        &  0.0678 &  0.8605\\
    \hline
    2    & 0.12 &  $48^3\times 64$    & $985\times 16$&   0.00184       &   0.0507         &0.0527  & 0.643\\
      \hline
    3   & 0.088 & $64^3\times 96$    & $620\times 8$&  0.00120       &  0.0363         &   0.036  &  0.433, 0.683, 0.8\\
      \hline
      \hline
    4    & 0.15 &  $16^3\times 48$    & $1020\times 16$&     0.013        &  0.065        &  0.0705 &  0.888\\
    \hline
    5    &  0.12 & $24^3\times 64$   & $1053\times 16$ &     0.0102        &  0.0509        &  0.0545  & 0.664, 0.8, 0.9\\
    \hline
    6    & 0.09 & $32^3\times 96$  & $499\times 16$  &   0.0074       &   0.037        &  0.0376  &  0.449, 0.566, 0.683, 0.8\\
    \hline
    7   & 0.059 & $48^3\times 144$    & $413\times 8$&  0.0048       &  0.024         & 0.0234 &  0.274, 0.45, 0.6, 0.8\\
    \hline
    8   & 0.044 & $64^3\times 192$    & $375\times 4$&  0.00316       &  0.0158        & 0.0165 &  0.194, 0.45, 0.6, 0.8\\
    \hline
    \hline
    \end{tabular}
  \end{center}
  \label{tab:ensembles}
\end{table}
$B\to K\ell\bar{\ell}$ decays involving $b\to s$ flavour changing neutral currents are highly suppressed loop processes in the standard model (SM) (see Figure~\ref{fig:BKfeyn}) and are good places to look for new physics. The increasing quantity of experimental data being collected demands better theoretical bounds on SM quantities to reveal new physics.

The uncertainties in current SM predictions are dominated by the hadronic form factors, which previously used the non-relativistic (NRQCD) formalism for the $b$ quark~\cite{Bouchard:2013pna}, or the Fermilab~(\cite{ElKhadra:1996mp}) interpretation~\cite{Bailey:2015dka}. Both of these methods rely on $\mathcal{O}(\alpha_s)$ perturbation theory to match the lattice weak current operators to their continuum counterparts.

In this proceedings, we present the first fully relativistic calculation, using the HISQ formalism for all valence quarks and working on MILC $N_f=2+1+1$ gluon field ensembles that include HISQ quarks in the sea~\cite{MILC:2010pul,MILC:2012znn}. This eliminates the aforementioned perturbative matching, and associated errors. 

Here, we summarise the content of~\cite{BK}, which covers the lattice calculation of the $B\to K$ form factors, and~\cite{BKpheno}, which contains the SM predictions arising from these new form factor determinations. See these references for more detail. 

\section{Lattice calculation details}
\subsection{Heavy HISQ}
The calculation detailed in~\cite{BK} uses the heavy-HISQ method. Because working at the physical $b$ mass requires extremely fine (and so costly) lattice spacings, $a$, to achieve small discretisation effects, we instead work with a set of  generic heavy quarks of mass $m_h$, which result in heavy mesons of mass $M_H$, covering a range of values. In Table~\ref{tab:ensembles}, we detail each of the 8 MILC ensembles that we use in this work. We use a variety of heavy masses on each ensemble, generally in the range $am_c\leq am_h\leq 0.8$. We must then perform an extrapolation to the physical $b$ mass $m_h\to m_b$ ($M_H\to M_B$). This method ultimately allows us to calculate the form factors across the full physical $q^2$ range, and evaluate our results at any mass between $M_D$ and $M_B$.  
\subsection{Form factor calculation}
We wish to calculate the scalar, vector and tensor form factors, $f_0(q^2)$, $f_+(q^2)$ and $f_T(q^2)$ as functions of the 4-momentum transfer $q^2=(M_H-E_K)^2-(\vec{p}_K)^2$, where we work in the rest frame of the parent $H$ meson. We can construct form factors using the following expressions,
\begin{equation}\label{Eq:vec}
  \begin{split}
    Z_V\bra{K}V^{\mu}_{\mathrm{latt}}\ket{\hat{H}}&= f_+(q^2)\Big(p_{H}^{\mu}+p_{K}^{\mu}-\frac{M_{H}^2-M_{K}^2}{q^2}q^{\mu}\Big)+f_0(q^2)\frac{M_{H}^2-M_{K}^2}{q^2}q^{\mu},\\
    \bra{K}S_{\mathrm{latt}}\ket{H}&=\frac{M_{H}^2-M_{K}^2}{m_h-m_s}f_0(q^2),\\
    Z_T(\mu)\bra{\hat{K}}T^{k0}_{\mathrm{latt}}\ket{\hat{H}}&=\frac{2iM_{H}p_K^k}{M_H+M_K}f_T(q^2,\mu),
  \end{split}
\end{equation}
where the scalar, vector and tensor matrix elements are extracted from multi-exponential Bayesian fits to three-point correlation functions calculated on the lattice (see~\cite{BK} for details). The vector current is normalised via $Z_V$ using the PCVC relation~\cite{Na:2010uf, Koponen:2013tua}. The tensor current renormalisation factor $Z_T$ is calculated in~\cite{Hatton:2020vzp}, using RI-SMOM.

Applying twisted boundary conditions to the strange daughter quark for a range of different momenta gives us good coverage of the $q^2$ range on each ensemble.

\subsubsection{Extrapolating to the continuum}
Taking our form factor data points at different $am_h$, and $q^2$ values, we perform a modified $z$ expansion using the Bourreley-Caprini-Lellouch (BCL) parameterisation~\cite{Bourrely:2008za},
\begin{equation}\label{Eq:zexpansion}
  \begin{split}
    f_0(q^2)&=\frac{\mathcal{L}}{1-\frac{q^2}{M^2_{H_{s0}^{*}}}}\sum_{n=0}^{N-1}a_n^0z^n\\
    f_{+,T}(q^2)&=\frac{\mathcal{L}}{1-\frac{q^2}{M^2_{H_{s}^{*}}}}\sum_{n=0}^{N-1}a_n^{+,T}\Big(z^n-\frac{n}{N}(-1)^{n-N}z^N\Big),
  \end{split}
\end{equation}
where $z(q^2)=\frac{\sqrt{t_+-q^2}-\sqrt{t_+}}{\sqrt{t_+-q^2}+\sqrt{t_+}}$ and $\mathcal{L}$ is a hard pion chiral logarithm term~\cite{Bijnens:2010jg}. The coefficients,
\begin{equation}\label{Eq:an}
\begin{split}
  a_n^{0,+,T}=&\Big(\frac{M_D}{M_H}\Big)^{\zeta_n}\Big(1+\rho_n^{0,+,T}\log\Big(\frac{M_{H}}{M_{D}}\Big)\Big)\times(1+\mathcal{N}^{0,+,T}_n)\times\\
  &\sum^{N_{ijkl}-1}_{i,j,k,l=0}d_{ijkln}^{0,+,T}\Big(\frac{\Lambda_{\text{QCD}}}{M_{H}}\Big)^i\Big(\frac{am_h^{\text{val}}}{\pi}\Big)^{2j}\Big(\frac{a\Lambda_{\text{QCD}}}{\pi}\Big)^{2k}(x_{\pi}-x_{\pi}^{\mathrm{phys}})^{l}
\end{split}
\end{equation}
contain terms allowing for discretisation effects ($am_h^{\text{val}}/\pi$ and $a\Lambda_{\text{QCD}}/\pi$) and quark mistunings ($\mathcal{N}^{0,+,T}_n$ and $x_{\pi}=M^2_{\pi}/(4\pi f_{\pi})^2$) (see~\cite{BK}), as well as a Heavy Quark Effective Theory inspired fit to the heavy meson mass $M_H$. The coefficients $\rho_n^{0,+,T}$, $d_{ijkln}^{0,+,T}$ and $\zeta_0$ are tunable fit coefficients, with $\zeta_{n\neq0}=0$. This combined $q^2$ and $M_H$ fit allows us to extrapolate our results to the physical point; $a=0$ and $M_H=M_B$.
\begin{figure}
  \begin{center}
    \includegraphics[width=0.78\textwidth]{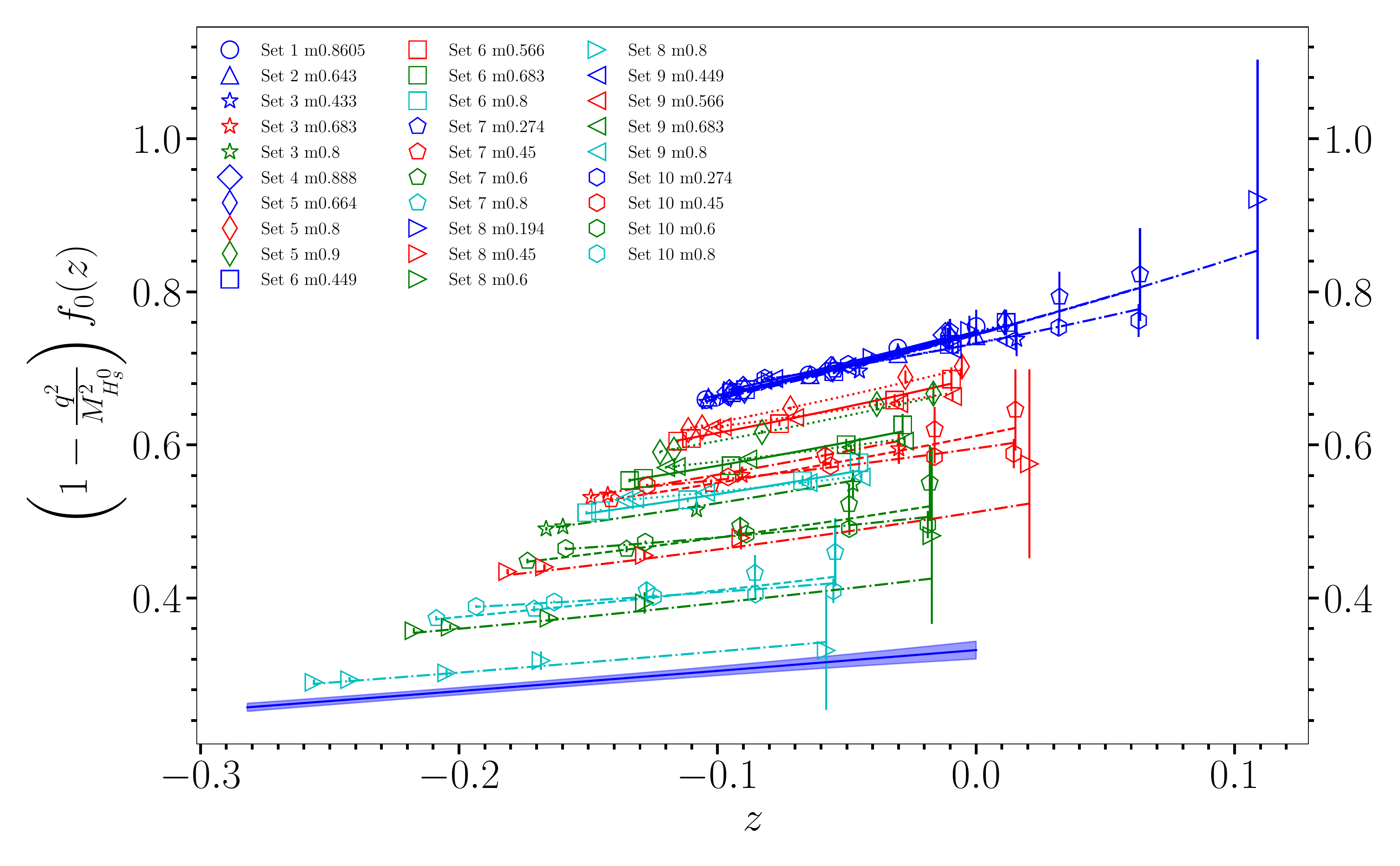}
    \includegraphics[width=0.78\textwidth]{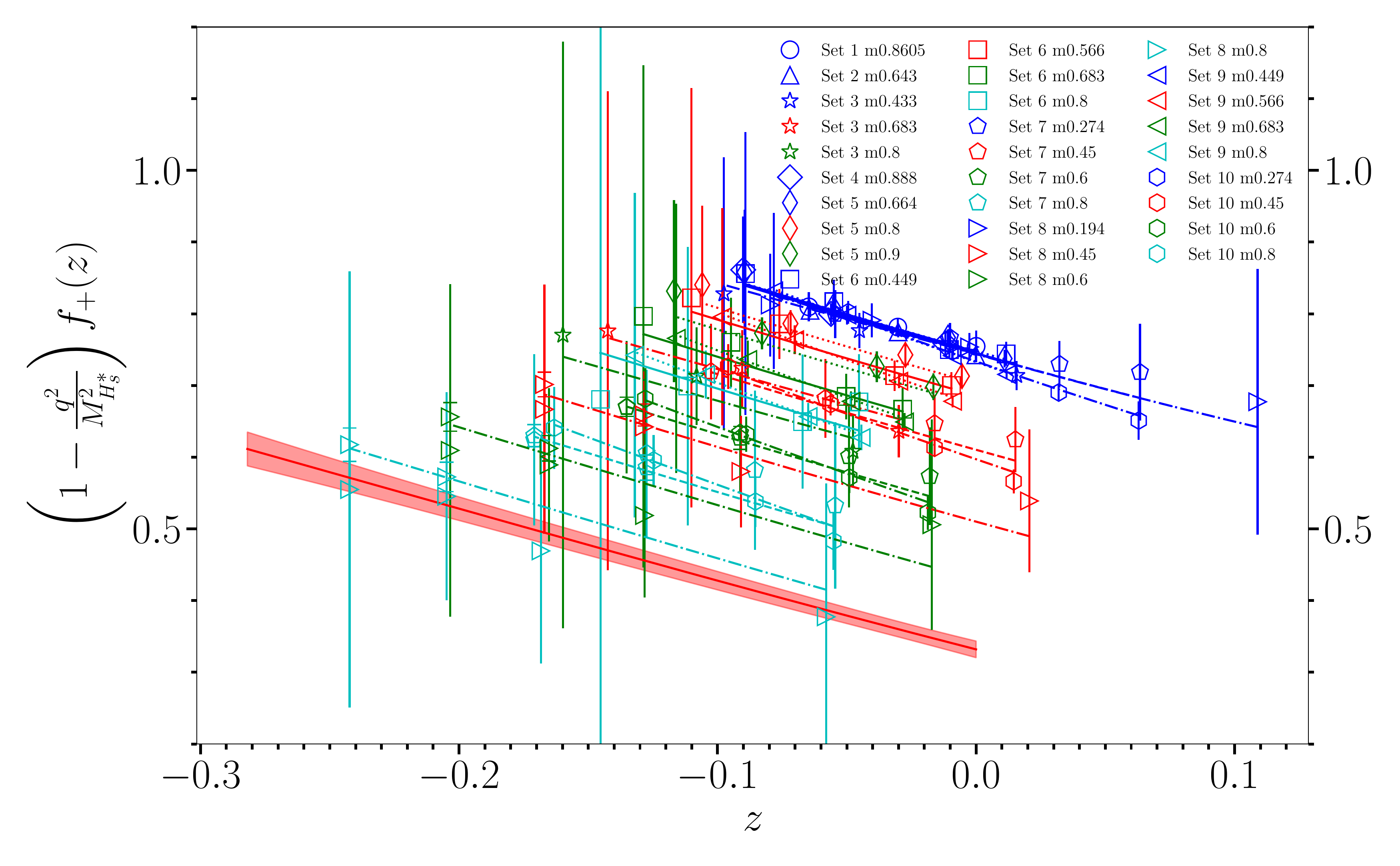}
    \includegraphics[width=0.78\textwidth]{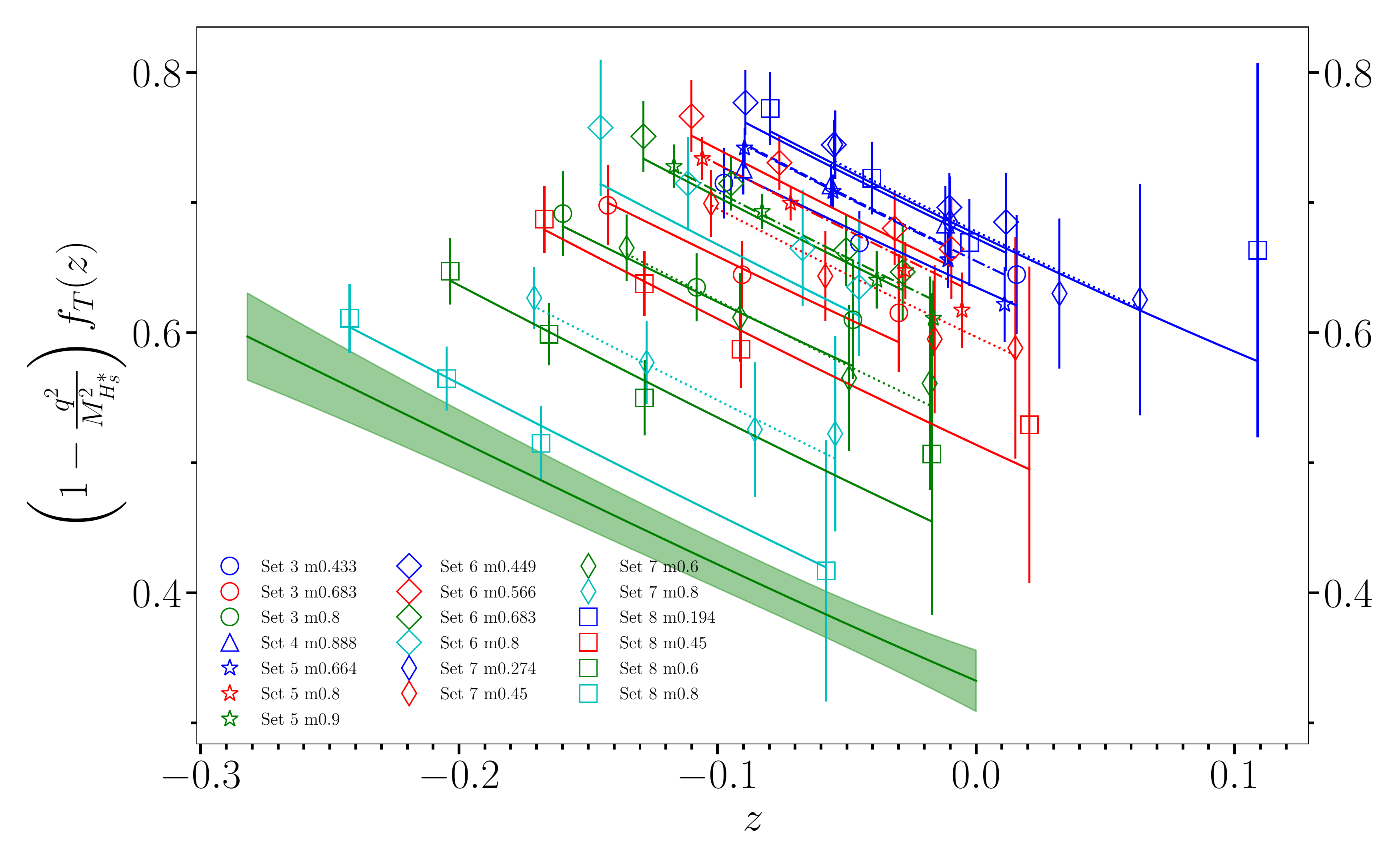}
    \caption{The scalar, vector and tensor form factors, with poles removed, in $z$ space. Data points on each ensemble are included as well as a solid band indicating the continuum result for $B\to K$.}
    \label{fig:ff_res}
  \end{center}
\end{figure}
\subsection{Results}
Figure~\ref{fig:ff_res} shows the continuum form factors alongside the data points on each ensemble, in $z$ space, with the pole removed. We see that $z$ space polynomials are very nearly linear, and that the fit is well behaved. Data on set 8, our finest ensemble, at $am_h=0.8$ (labelled $m0.8$), our highest mass, closely approaches the physical point, as we would expect, given that this is very close to the physical $B$ mass. The data points labelled as sets 9 and 10 are correlated $B_s\to\eta_s$ data on sets 6 and 7, taken from~\cite{Parrott:2020vbe} (see~\cite{BK}).
\begin{figure}
  \begin{center}
    \includegraphics[width=0.78\textwidth]{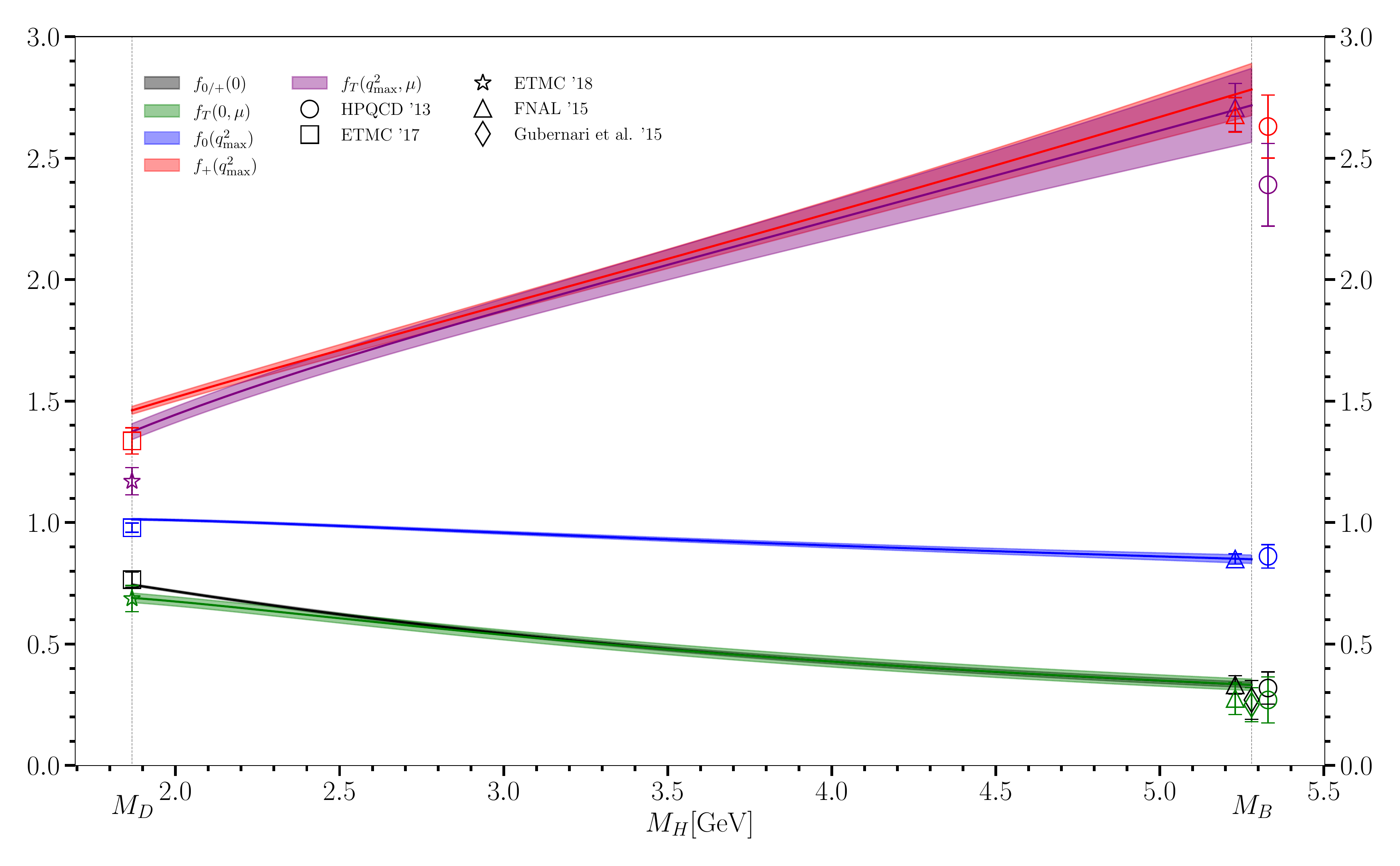}
    \caption{The scalar, vector and tensor form factors, at $q^2=0$ and $q^2_{\mathrm{max}}$, plotted against $M_H$. Other results~\cite{Riggio:2017zwh,Lubicz:2018rfs,Bouchard:2013pna,Bailey:2015dka,Gubernari:2018wyi} are included for comparison. For a discussion of the scale, $\mu$, see~\cite{BK}.}
    \label{fig:ff_in_mH}
  \end{center}
\end{figure}

Because our heavy-HISQ approach requires a fit in the heavy mass, with $am_c\leq am_h<am_b$ we are able to evaluate the form factors at any mass between $M_D$ and $M_B$. Figure~\ref{fig:ff_in_mH} shows the scalar, vector and tensor form factors, at extremal $q^2$ values (note $f_+(0)=f_0(0)$), plotted against $M_H$. We can see that we agree well with previous work~\cite{Riggio:2017zwh,Lubicz:2018rfs,Bouchard:2013pna,Bailey:2015dka,Gubernari:2018wyi}, across the $M_H$ and $q^2$ range, with the exception of $f_T(q^2_{\mathrm{max}},2~\mathrm{GeV})$ and $f_+(q^2_{\mathrm{max}})$ from ETMC~\cite{Riggio:2017zwh,Lubicz:2018rfs}. Our uncertainties at $q^2=0$ are improved by a factor of three over previous work.  
\section{Phenomenology}
\subsection{$B\to K\ell^+\ell^-$}
We can write the SM differential decay rate for $B\to K\ell^+\ell^-$ in terms of $f_0$, $f_+$ and $f_T$, as well as the Wilson coefficients $W_i$. This lengthy expression is discussed in detail in~\cite{BKpheno}. Schematically,
\begin{equation}\label{eq:diff}
  \frac{d\Gamma^{B\to K\ell^+\ell^-}}{dq^2} = \mathcal{F}_1|F_P(f_0,f_+,W_i)|^2+\mathcal{F}_2f_+^2 + \mathcal{F}_3|F_V(f_+,f_T,W_i)|^2 + \mathcal{F}_4|f_+F_P^*(f_0,f_+,W_i)|, 
\end{equation}
where $\mathcal{F}_i$ are known functions of kinematic factors and $W_i$. This expression does not account for production of $c\bar{c}$ resonances, so is not valid in all regions of $q^2$. We can compare this differential decay rate (or equivalently the differential branching fraction $\mathcal{B}$, noting that $\mathcal{B}=\Gamma\tau_B$, for $B$ lifetime $\tau_B$), directly with experiment in different $q^2$ regions, to look for evidence of new physics.

\begin{figure}
  \begin{center}
    \includegraphics[width=0.48\textwidth]{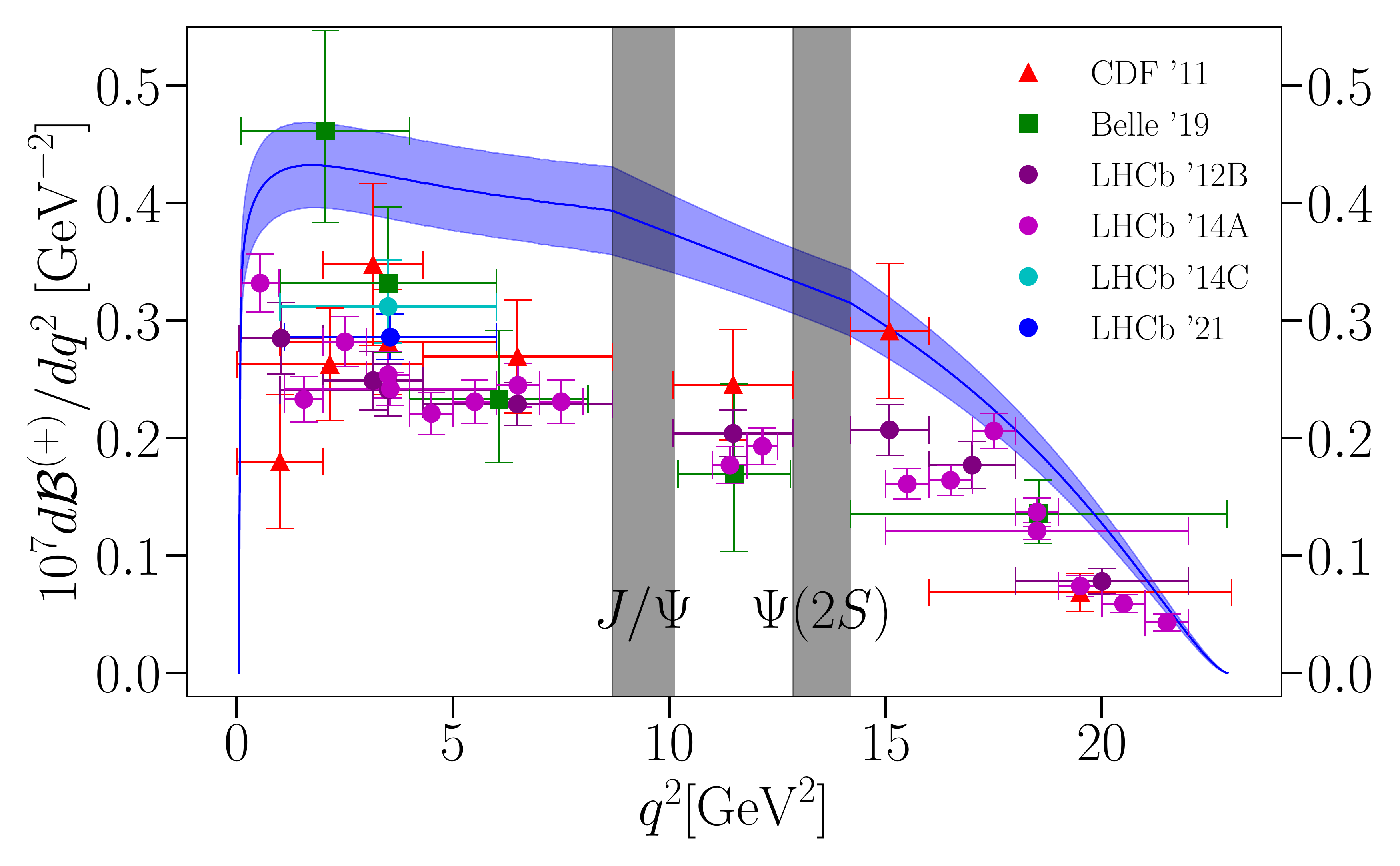}
    \includegraphics[width=0.48\textwidth]{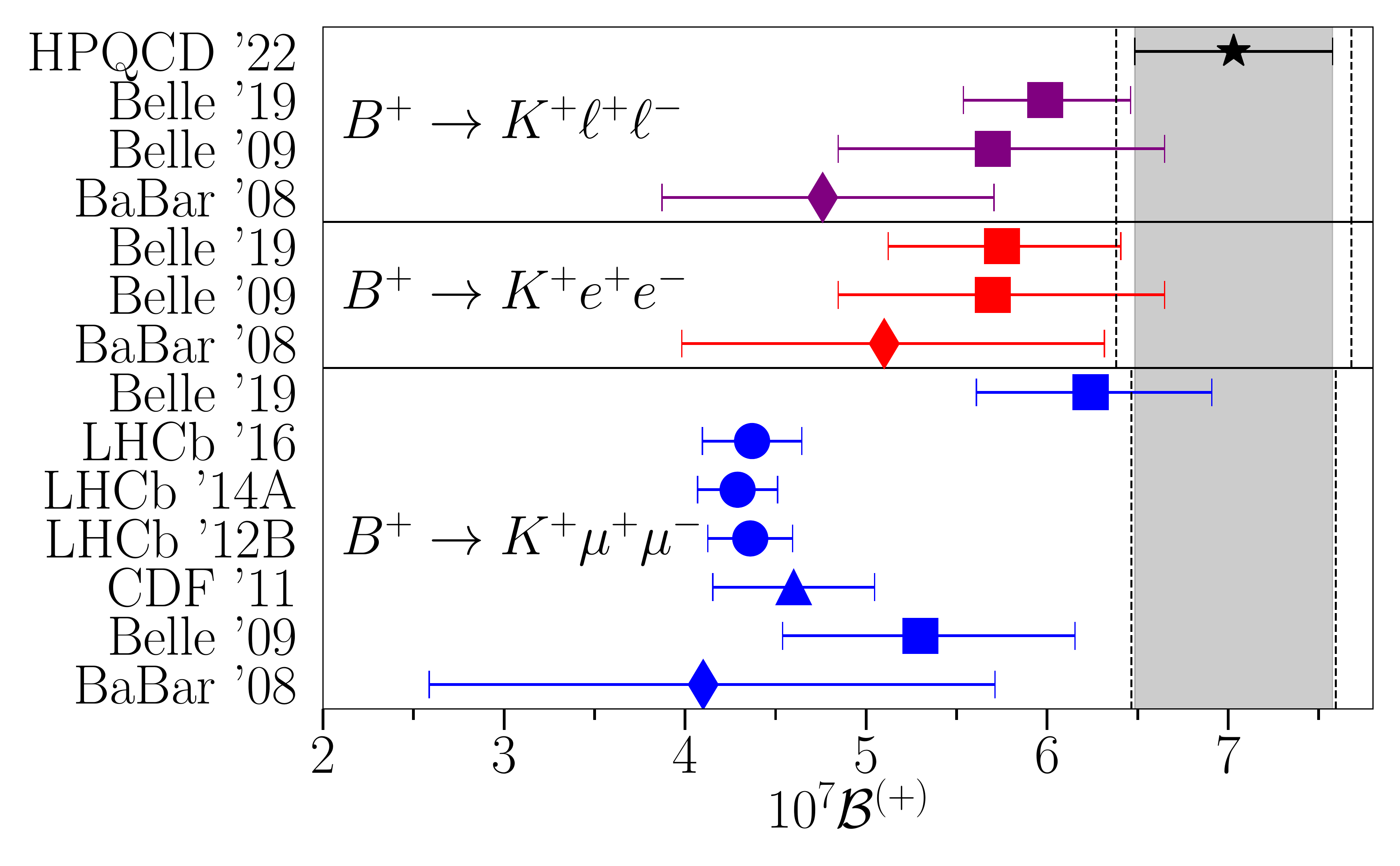}
    \caption{Left: our result for $d\mathcal{B}(B^+\to K^+e(\mu)^+e(\mu)^-)/dq^2$ (blue band), compared with experiment~\cite{Aaltonen:2011qs,Aaij:2012vr,Aaij:2014pli,Aaij:2014ora,BELLE:2019xld,Aaij:2021vac}. We linearly interpolate across both vetoed regions in grey, and the region between them. Right: the same result, but integrated across the full $q^2$ range, compared with experiment~\cite{Aubert:2008ps,Wei:2009zv,Aaij:2014pli,Aaij:2012vr,Aaltonen:2011qs,LHCb:2016due,BELLE:2019xld}. Our value is carried down the page in grey. The dotted lines indicate additional QED uncertainty (see~\cite{BKpheno}).}
    \label{fig:diff}
  \end{center}
\end{figure}
\begin{figure}
  \begin{center}
    \includegraphics[width=0.6\textwidth]{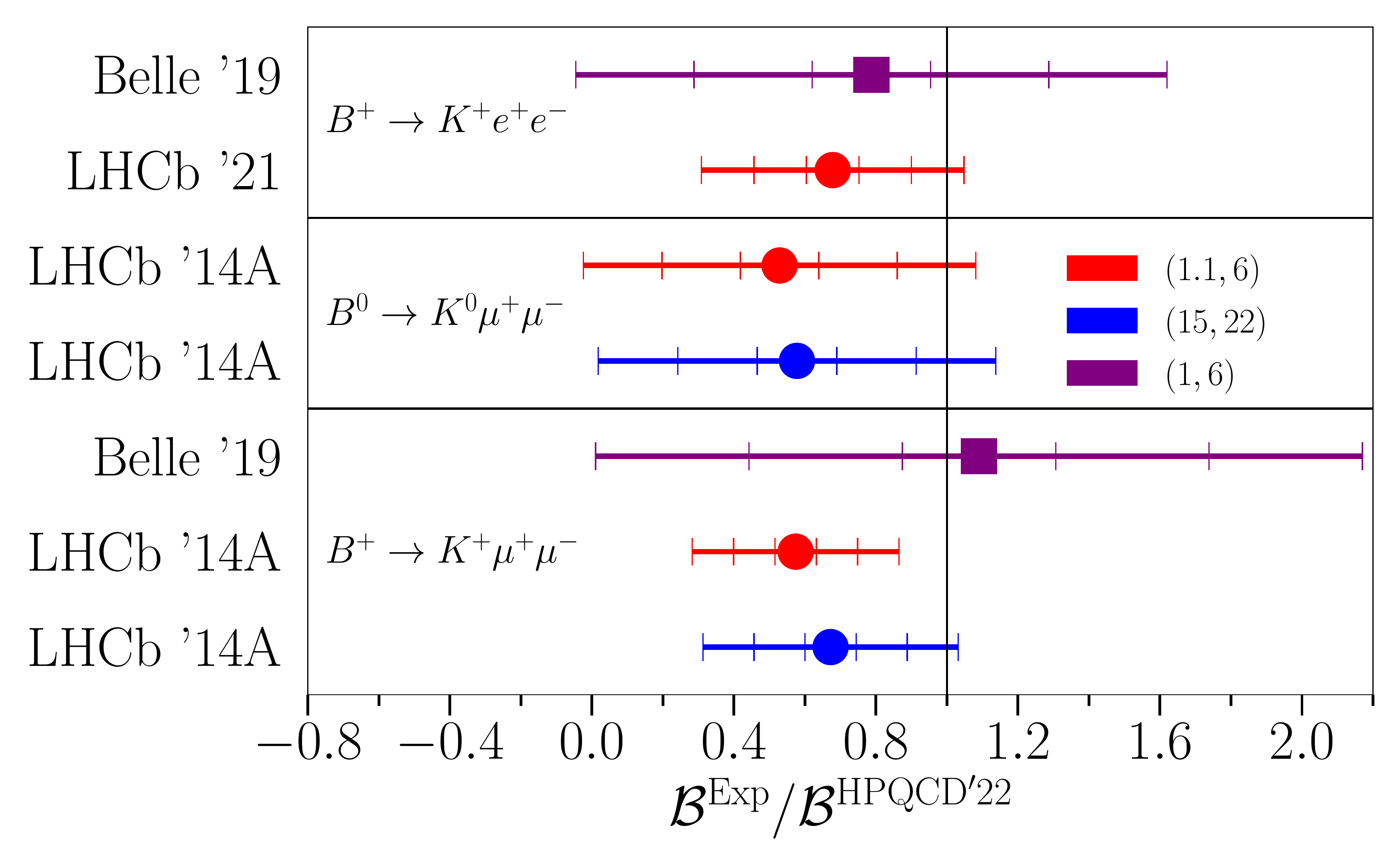}
    \caption{The ratio of experimental branching fractions~\cite{Aaij:2014pli,BELLE:2019xld,Aaij:2021vac} to our result, over some theoretically clean $q^2$ bins. (1(.1)-6, and 15-22$\mathrm{GeV}^2$ respectively). Caps on error bars indicate 1, 3 and 5 $\sigma$. }
    \label{fig:headline}
  \end{center}
\end{figure}

The left of Figure~\ref{fig:diff} shows our result for $d\mathcal{B}(B^+\to K^+e(\mu)^+e(\mu)^-)/dq^2$, across the full $q^2$ range, as well as data from experiment. Grey bands indicated vetoed regions where $c\bar{c}$ resonances dominate and mean that Eq.~\eqref{eq:diff} is no longer valid. We see that our result lies considerably above experiment, particularly at low $q^2$. On the right of the same figure, we show the result when integrated across the full $q^2$ range, interpolating across the vetoed regions, as is done in experiment.

Focusing specifically on the most recent experimental results in $q^2$ bins away from the $c\bar{c}$ resonances, we plot experimental branching fractions, divided by our result, in Figure~\ref{fig:headline}. Error bars have caps at 1, 3 and 5 $\sigma$. We see that we are in $3-5\sigma$ tension with LHCb~\cite{Aaij:2014pli} results in these theoretically clean regions of $q^2$.  
\subsection{$B\to K\nu\bar{\nu}$}
\begin{figure}
  \begin{center}
    \includegraphics[width=0.48\textwidth]{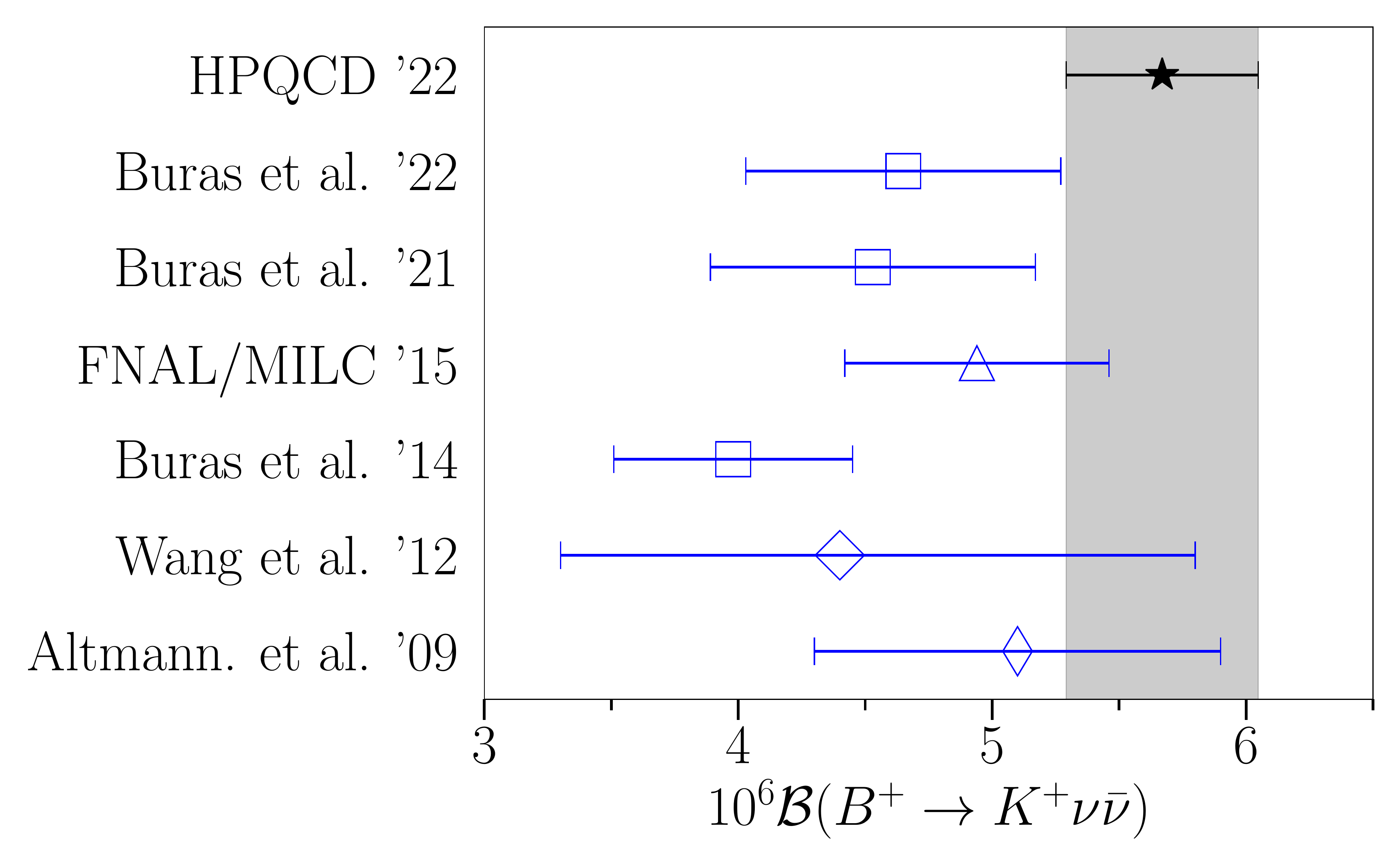}
    \includegraphics[width=0.48\textwidth]{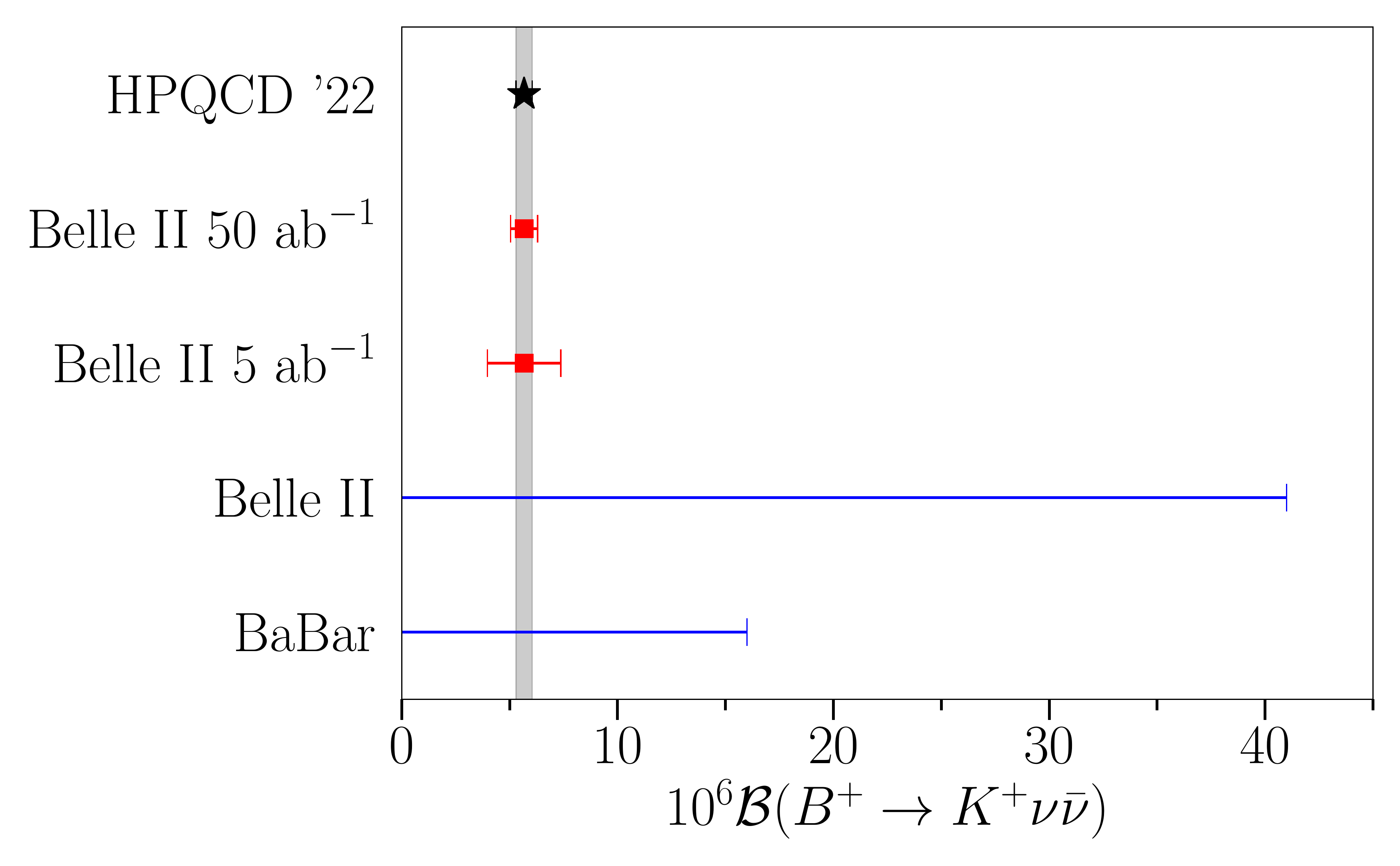}
    \caption{The branching fraction $B^+\to K^+\nu\bar{\nu}$ compared with other theoretical work~\cite{Altmannshofer:2009ma,Kamenik:2009kc,Wang:2012ab,Buras:2014fpa,Du:2015tda,Buras:2021nns,Buras:2022wpw} (left) and experimental bounds (right)~\cite{Lees:2013kla,Belle-II:2021rof}. Experimental bounds in blue are current 90\% confidence limits. The red points indicate the expected future uncertainty from Belle II at 5 and 50 $\mathrm{ab}^{-1}$~\cite{Halder:2021sgd}, centred on our result.}
    \label{fig:nu}
  \end{center}
\end{figure}
A much cleaner channel, free from resonances, that we can also probe, is $B\to K\nu\bar{\nu}$, although experimental measurements are much less mature here. We can write the short distance (SD) differential branching fraction,
\begin{equation}
  \frac{d\mathcal{B}(B\to K\nu\bar{\nu})_{\text{SD}}}{dq^2} = \frac{(\eta_{\mathrm{EW}}G_F)^2\alpha_{\mathrm{EW}}^2X_t^2}{32\pi^5\sin^4\theta_W}\tau_B|V_{tb}V^*_{ts}|^2|\vec{p}_K|^3f^2_+(q^2),
\end{equation}
where the values used are given in~\cite{BKpheno}. A well determined $~10\%$ correction is added to account for long distance effects in the case $B^+\to K^+$ (see~\cite{BKpheno}).

In Figure~\ref{fig:nu}, we present our result for the branching fraction, as well as other theoretical and experimental work. Our number is slightly larger and more precise than previous theoretical work, and well within the current experimental confidence bounds. We have similar uncertainty to that expected from Belle II with $50\mathrm{ab}^{-1}$ of data~\cite{Halder:2021sgd}.  
\section{Conclusions}
We present the first fully relativistic determination of the scalar, vector and tensor form factors for $B\to K$ decays. We have good agreement with previous determinations, and improved uncertainty at low $q^2$.

Using these form factors, we can calculate the SM branching fraction for $B\to Ke(\mu)^+e(\mu)^-$ in $q^2$ bins which are free from $c\bar{c}$ resonances and measured experimentally. We find tensions above $3\sigma$ in the ratio of our result with LHCb~\cite{Aaij:2014pli}, and above $5\sigma$ in one instance for the low $q^2$ bin.

We also determine the theoretically clean branching fraction $B\to K\nu\bar{\nu}$, finding a value which is more precise than previous experimental work. Our result has an uncertainty below 10\%, which is commensurate with the uncertainty expected from Belle II in future, with $50\mathrm{ab}^{-1}$ of data~\cite{Halder:2021sgd}.
\section{Acknowledgements}
We are grateful to the MILC collaboration for the use of  their  configurations  and  their  code, which we use to generate quark propagators and construct correlators. We would also like to thank  L. Cooper, J. Harrison, D. Hatton and G. P. Lepage for useful discussions and B. Chakraborty, J. Koponen and A. T. Lytle for generating propagators/correlators in previous projects that we could make use of here. Computing was done on the Cambridge Service for Data Driven Discovery (CSD3) supercomputer, part of which is operated by the University of Cambridge Research Computing Service on behalf of the UK Science and Technology Facilities Council (STFC) DiRAC HPC Facility. The DiRAC component of CSD3 was funded by BEIS via STFC capital grants and is operated by STFC operations grants. We are grateful to the CSD3 support staff for assistance. Funding for this work came from STFC. 
\bibliographystyle{JHEP}
\bibliography{BKpaper}
\end{document}